\documentclass[fleqn,usenatbib]{mnras}

\usepackage[T1]{fontenc}

\DeclareRobustCommand{\VAN}[3]{#2}
\let\VANthebibliography\thebibliography
\def\thebibliography{\DeclareRobustCommand{\VAN}[3]{##3}\VANthebibliography}

\usepackage{graphicx}	
\usepackage{amsmath}	
\usepackage{amssymb}	

\usepackage{newtxtext,newtxmath}

\title[Gravitational waves from spin-up and spin-down events]{Gravitational waves from small spin-up and spin-down events of neutron stars}

\author[G. Yim \& D. I. Jones]{
Garvin Yim\thanks{E-mail: g.yim@soton.ac.uk} and
D. I. Jones\thanks{E-mail: d.i.jones@soton.ac.uk} 
\\
Mathematical Sciences and STAG Research Centre, University of Southampton, Southampton SO17 1BJ, UK
}

\date{Accepted XXX. Received YYY; in original form ZZZ}

\pubyear{2022}

\begin{document}
\label{firstpage}
\pagerange{\pageref{firstpage}--\pageref{lastpage}}
\maketitle

\begin{abstract}
It was recently reported that there exists a population of ``glitch candidates'' and ``anti-glitch candidates'' which are effectively small spin-ups and spin-downs of a neutron star with magnitudes smaller than those seen in typical glitches. The physical origin of these small events is not yet understood. In this paper, we outline a model that can account for the changes in spin, and crucially, is independently testable with gravitational wave observations. In brief, the model posits that small spin-up/spin-down events are caused by the excitation and decay of non-axisymmetric $f$-modes which radiate angular momentum away in a burst-like way as gravitational waves. The model takes the change in spin frequency as an input and outputs the initial mode amplitude and the signal-to-noise ratio achievable from gravitational wave detectors. We find that the model presented here will become falsifiable once 3rd generation gravitational wave detectors, like the Einstein Telescope and Cosmic Explorer, begin taking data.
\end{abstract}




\section{Introduction}
\label{section_introduction}


Neutron stars (NSs) are extremely stable rotators, with spin frequencies that are generally observed to decrease with time; see \citet{l_g-s_12} for a review. However, they exhibit small deviations from smooth spin down. These deviations consist of relatively large steps in spin frequency, known as \textit{glitches}, and smaller deviations often known as \textit{timing noise}, not normally resolved into individual events. \cite{espinozaetal2014, espinozaetal2021} have added to this picture by identifying a population of small timing events, that could be interpreted as spin-ups, smaller than previously resolved glitches, which clearly form a separate population of events from glitches. They also found evidence for similar small spin-down events. They dubbed these two sets of events as \textit{glitch candidates} (GCs) and \textit{anti-glitch candidates} (AGCs), respectively.

As defined by \cite{espinozaetal2014, espinozaetal2021}, a GC is a timing event where a neutron star instantaneously increases its spin frequency, $\Delta \nu > 0$, and has a simultaneous decrease to the time derivative of the spin frequency, $\Delta \dot{\nu} \le 0$, effectively mimicking a small glitch (see \cite{espinozaetal2011, yuetal2013, loweretal2021, basuetal2022} for more on glitches). For an AGC, the opposite applies ($\Delta \nu < 0$, $\Delta \dot{\nu} \ge 0$).  Unlike most glitches, the recovery is not resolvable with current ($\sim$daily) observational cadences and so changes to $\nu$ and $\dot{\nu}$ are treated as step-like. The analysis of \cite{espinozaetal2014, espinozaetal2021} showed that these small events are significant enough to be distinguished from detector noise so make up a population of events that should have a physical explanation. However, the production mechanism need not be the same as glitches since GCs/AGCs can have either sign of $\Delta \nu$ and are observed to form a separate population to glitches on a $\Delta \nu - \Delta \dot{\nu}$ plot (e.g.~see Fig.~2 of \cite{espinozaetal2014} or Fig.~2 of \cite{espinozaetal2021}). 

In fact, this is not the first time small events of this sort have been reported. \cite{cordesDowns1985} and \cite{cordesDownsKrause-Polstroff1988} also found evidence of such small events, and concluded that typical glitch models like starquakes and vortex unpinning (see also \cite{alparNandkumarPines1986}) could not be responsible, particularly for AGCs. There is also the question of whether GCs and AGCs contribute towards timing noise, for which there already exists several models \citep[e.g.][]{chengetal1988, lyneetal2010, jones2012, melatosLink2014}. However, during early times, many leading models were ruled out \citep{cordesGreenstein1981}, leaving the true timing noise mechanism still uncertain. In this paper, we take a step towards resolving this by suggesting a physical explanation for the observed GC/AGC events.  As well as potentially explaining these events, our model predicts a calculable level of gravitational wave (GW) emission, which offers an independent test for the model.

The study of GWs has been accelerating over the last few years thanks to the first detection of a GW signal in 2015 from the coalescence of two black holes \citep{LVC2016firstdetection}.  The detections of further GWs from NS-NS and NS-black hole binaries has added to the success story for GWs \citep{LVK2021GWTC3}. All existing detections fall under the ``compact binary coalescence'' category.  One of the main goals over the next decade is to detect other types of GWs, of a continuous, burst or stochastic nature.

Modelling efforts have already suggested possible sources of GW bursts, e.g. \citet{LVK2020prospectsreview, LVK2021allskyshortburstsearch}, including several from NS oscillations.  These include the excitation of stellar oscillations after birth \citep{ferrariMiniuttiPons2003}, after magnetar flares \citep{ioka2001, corsiOwen2011, LVK2022magnetarbursts}, after pulsar glitches \citep[e.g.][]{keerJones2015, hoetal2020} or after a binary NS coalescence where the remnant NS survives sufficiently long \citep[e.g.][]{clarketal2016}.  We add to this list by proposing a model in which GC/AGC events represent the sudden excitation of a non-axisymmetric mode of oscillation of the NS, with an accompanying short ($< 1~\text{s}$)  burst of GWs.

NS oscillations can be grouped into different types depending on the physics that gives rise to them. $f$-modes, named so because they are the fundamental modes, are modes that cause the entire shape of the NS to deform, so that there are no radial nodes within the NS. $f$-modes are some of the most efficient oscillations to emit GWs \citep{mcdermottVanHornHansen1988, anderssonKokkotas1998} so we choose them as the modes we excite in our model. Furthermore, NS oscillations are of particular interest because, like with astroseismology or helioseismology, seismology of NSs can reveal information about the elusive interior \citep{anderssonKokkotas1998, andersson2021}. The work presented here does not address the interesting question of how different interiors may affect our results, but this could (and should) be done at a later date. For now, we will use the simplest model possible to achieve analytic results and provide a proof of concept. 

The main assumptions and concept of the model are as follows. We assume that the total angular momentum of a slowly rotating isolated NS can be broken down into two parts, the background and the mode. This is typically what is done in $r$-mode analyses \citep{owenetal1998, hoLai2000, levinUshomirsky2001}. We also assume the electromagnetic timing of the NS is tied to the background. Then, in an isolated system, the sudden excitation of a non-axisymmetric $f$-mode, which carries angular momentum, induces a small change in the rotation of the background of opposite sign, in order to conserve angular momentum. The $f$-mode radiates away its angular momentum to infinity as GWs, leaving a net change to the background, manifesting observationally as a small positive or negative change in the spin frequency. The details of this calculation are covered in \cite{yimJones2022} and will be summarised in Section~\ref{section_the_model}. The model presented is independent of how the modes are excited (e.g. starquakes, vortex unpinning) and applies to NSs that rotate much slower than their Keplerian break-up frequency, as this is what was assumed in \cite{yimJones2022}. This work presented here supplements our previous work by taking a closer look at the GWs given off from such a model.

The paper is organised as follows. In Section~\ref{section_the_model}, we introduce and develop the equations that make up the model. By the end of that section, we will have an expression for the change in spin frequency as a function of the initial mode amplitude. In Section~\ref{section_connecting_to_gravitational_waves}, we use this mode amplitude to find the corresponding GW strain which we then use to find an expression for the GW signal-to-noise ratio (SNR). Following this, we use the results of Section~\ref{section_the_model} to write the SNR as a function of the observed change in spin frequency. In Section~\ref{section_applying_to_data}, we apply the calculations to GC and AGC data from \citet{espinozaetal2014, espinozaetal2021} and also explore a speculative scenario where glitches themselves are caused by the same mechanism. In Section~\ref{section_energetics}, we evaluate the energy budget required to power the modes and in Section~\ref{section_summary}, we provide a summary and discussion of our findings.

\section{The model}
\label{section_the_model}

We will model the star in a very simple way, as a uniform density self-gravitating ball, spinning with angular frequency $\Omega~(= 2 \pi \nu)$. We imagine that as the star spins down, strains builds up, due perhaps to some combination of the deformation of the elastic crust \citep{baymPines1971} and/or the Magnus force on any pinned superfluid \citep{jones2010}. Our central assumption is that these strains are eventually relieved in an impulsive way, and that at least part of the star's response is the excitation of a non-axisymmetric $f$-mode. That elasticity and pinned superfluidity can induce oscillations is clear, and has been explored in simplified form already; see \citet{keerJones2015} and \citet{sideryPassamontiAndersson2010}.

By modelling the star as a simple fluid, we forgo any possibility of describing this excitation process.  However, it has been shown that elasticity \citep{krugerHoAndersson2015, floresHallJaikumar2017} and superfluidity \citep{gualtierietal2014} have little effect on the properties of $f$-modes, i.e.~on their damping times and mode frequencies. This therefore allows us to proceed in describing the star's subsequent evolution without incorporating these complicating but very small effects, simplifying the equations greatly and allowing for a fully analytic treatment, making use of the results of  \citet{yimJones2022}.  In this way, we are completely agnostic to the precise mode excitation mechanism; we simply assume such a mechanism exists, and explore its consequences.

Most of the equations for stellar oscillations and their GW emission in this section come from our earlier work \citep{yimJones2022} which assumes the NS is uniformly-dense, incompressible and (initially) non-rotating, which is a good approximation for NSs rotating much slower than their Keplerian break-up frequency.  Rotational corrections to $f$-modes will have leading order terms on the order of $\mathcal{O}(\nu / f_\text{K})$, where $f_\text{K}$ is the Kepler frequency and is around 1.0~-~1.2~kHz for a $1.4~\text{M}_\odot$ NS \citep{haenseletal2009}. We refrain from doing the full rotation $f$-mode calculation as it will have little influence on our final results. These modes of non-rotating uniform density stars are often known as the \textit{Kelvin modes}.

Following the convention of \cite{yimJones2022}, we consider modes with an oscillatory dependence on time $t$ and azimuthal angle $\phi$ given by $e^{i(m\phi + \omega t)}$, where $l$ and $m$ are the spherical harmonic numbers and $\omega$ is the (inertial frame) mode frequency. For a given $l$, $m$ ranges from $-l$ to $l$ in integer steps. In the analysis presented here, we will focus on $l=2$ since the (mass) quadrupole is the strongest emitter of GWs \citep{thorne1980}, but the inclusion of $l>2$ can be trivially incorporated. Also, we consider only $m = \pm 2$, as such $m \neq 0$ modes carry angular momentum, which is an important ingredient of our model.  There is an open question of how a local defect, e.g. a crack in the crust or an unpinning event localised to the inner crust, can grow to cause a global $l=2$ deformation, but we leave this to be answered in future studies. 

In any case, assuming the excitation of a non-axisymmetric mode can occur, exciting a mode with angular momentum $\delta J$ will change the background's angular momentum by $-\delta J$ by the conservation of angular momentum. The non-axisymmetric mode can be thought of as some deformation pattern that propagates in the positive ($\delta J > 0$, $m < 0$) or negative ($\delta J < 0$, $m > 0$) mathematical sense, at pattern speed $\omega_\text{p} = - \omega / m$. 

Since the mode causes a time-varying mass quadrupole moment, GWs will be emitted which causes the mode to decay. One might assume that a mode with angular momentum $\delta J$ can only release up to $\delta J$ in GW emission, but after carefully taking both energy and angular momentum into account, it has been shown that the mode actually emits $2 \delta J$ as GWs by the time it has fully decayed \citep{yimJones2022}. This has the effect of causing the background angular momentum to change by a further $- \delta J$. The net result of the mode excitation and decay is that $2 \delta J$ of angular momentum is emitted as GWs and the NS background gains $-2 \delta J$ of angular momentum. It should be noted that we are proposing that the excitation and decay of a non-axisymmetric mode is the cause of a spin-up/spin-down event, and not the other way around. 

From the above logic, we can see by straightforward angular momentum conservation
\begin{align}
\label{glitch_size_dJ}
I \Delta \Omega = - 2 \delta J(0)~~\rightarrow~~\frac{\Delta \Omega}{\Omega} = - \frac{2 \delta J(0)}{I\Omega}~,
\end{align}
where $\delta J(0)$ is the angular momentum given to the mode at $t=0$ (corresponding to when the spin-up/spin-down event occurs) and where we have assumed no net change in the background's moment of inertia ($=I$) during the excitation and decay of the mode.

In \cite{yimJones2022}, we were able to link the Kelvin mode angular momentum $\delta J$ to the mode amplitude $\alpha_{2,m}$ ($\sim \Delta r/R \ll 1$, where $\Delta r$ is the radial displacement of the NS surface), which is given by
\begin{align}
\label{delta_J}
\delta J(t) = -\frac{1}{2} m \alpha_{2,m}^2(t) \bar\rho \omega_2 R^5~,
\end{align}
where $m$ is the azimuthal spherical harmonic number ($-2 \le m \le 2$), $\bar{\rho}$ is the (uniform) mass density, $R$ is the NS radius and $\omega_{2}$ is the $l=2$ (Kelvin) mode angular frequency, given by
\begin{equation}
\omega_{2}^2 = \frac{16 \pi G \bar{\rho}}{15} = \frac{4GM}{5R^3}~,
\end{equation}
which has typical values of $f = \omega_{2} / 2\pi \approx 2~\text{kHz}$ for a canonical NS with $M = 1.4~\text{M}_\odot$ and $R = 10~\text{km}$. Note that for $m < 0$, we get $\delta J > 0$ which indicates a mode propagating in the positive mathematical sense and vice versa for $m > 0$. 

Once excited, the mode will decay due to the emission of GWs, giving an exponentially decaying time dependence 
\begin{align}
\label{alpha_decay}
\alpha_{2,m}(t) = \alpha_{2,m}(0) e^{-\frac{t}{\tau}}~,
\end{align}
where $\tau$ is the mode damping time-scale, calculated to be
\begin{align}
\label{tau_phys}
\tau = \frac{10c^5}{\omega_2^6 R^5} = \frac{625}{32} \frac{c^5}{G^3} \frac{R^4}{M^3}~;
\end{align}
see \cite{yimJones2022}.  Putting in values for a canonical NS gives $\tau \approx 0.07~\text{s}$. As the duration of GWs emitted will be of the order of the mode damping time-scale, it is clear that the GWs will be emitted as a burst, lasting $\lesssim 0.1~\text{s}$. Since this time-scale is so short, any change in the angular frequency would appear step-like, and this is indeed what is seen for GCs and AGCs in \citet{espinozaetal2014, espinozaetal2021}. Furthermore, this model predicts the observation of a two-step change if an observer's telescope can resolve times shorter than $\tau$, with the first step being due to mode excitation and the second due to mode decay. Such resolution is highly unlikely though as $\tau$ is already about the same as one period for a slowly-rotating NS.

Eliminating $\delta J(0)$ from equation~(\ref{glitch_size_dJ}) by using equation~(\ref{delta_J}) and using $I = \frac{2}{5} M R^2$, one finds
\begin{align}
\label{glitch_size_alpha}
\frac{\Delta \Omega}{\Omega} \approx \frac{15}{8 \pi} m \alpha_{2,m}^2(0) \frac{ \omega_{2}}{\Omega} ~.
\end{align}
We note that for $m > 0$, which represents a mode propagating in the negative sense, we get a positive change to the angular frequency, i.e.~a GC. For $m < 0$, which represents a mode propagating in the positive sense, we get a negative change to the angular frequency, i.e.~an  AGC. For a fixed $m$ and $\omega_{2}~(\propto \sqrt{\bar{\rho}})$, the only free parameter left is the initial mode amplitude which controls how much the angular frequency changes. Or, from the other direction, observations of a given glitch size (or GW strain, see later) corresponds to a certain mode amplitude, if this model is to be believed.

\section{Connecting to gravitational waves}
\label{section_connecting_to_gravitational_waves}

We have just demonstrated how the excitation and decay of a non-axisymmetric $f$-mode can account for a change in angular frequency. Now, we will look at the associated GWs that will be emitted. We will provide predictions for the emitted GW signal (strain) $h(t)$ and use it to find an expression for the optimal SNR achievable.

\subsection{Gravitational wave strain}
\label{subsection_gravitational_wave_strain}
To begin, we will write down the generic form for the GW strain expected from decaying modes which, for $t \ge 0$, is
\begin{align}
\label{exponentially_decaying_sinusoid}
h(t) \equiv h_0(t) \cos\left[\Phi(t)\right] \equiv h_0(0) e^{-\frac{t}{\tau}}  \cos\left[\Phi(t)\right]~,
\end{align}
where $h_0(t)$ is the GW amplitude and $\Phi(t)$ is the GW phase. This form is expected as $f$-modes give rise to sinusoidal behaviour but then GW damping causes the sinusoid to exponentially decay, with associated envelope decay time-scale $\tau$, the same as the mode damping time-scale which is given by equation~(\ref{tau_phys}).

To find $h_0(0)$, we consider the rate of energy loss by the emission of GWs. This is calculated from the standard GW luminosity quadrupole formula and for the special case of $l=2$, $m = \pm 2$, it is
\begin{align}
\label{GW_luminosity_general}
\dot{E}_\text{GW} = \frac{1}{10} \frac{c^3}{G} \omega_{2}^2 d^2 h_0^2~,
\end{align} 
where $c$ is the speed of light, $G$ is the gravitational constant and $d$ is the distance to the GW source. The convention being used here is that $\dot{E}_\text{GW}$ is positive when energy is being lost from the NS system. This equation is in agreement with equation~(21) of \cite{owen2010}. 

In \citet{yimJones2022}, we found that the GW luminosity from $l=2$ $f$-modes, in terms of $\alpha_{2,m}$, is given by
\begin{align}
\dot{E}_\text{GW} = \frac{1}{5c^5} \alpha_{2,m}^2 \bar{\rho} \omega_{2}^8 R^{10}~,
\end{align}
so when we equate to equation~(\ref{GW_luminosity_general}) and rearrange, we find
\begin{align}
\label{GW_signal}
h_0(t) = \frac{4}{25} \sqrt{\frac{30}{\pi}} \alpha_{2, 2}(0) \frac{G^2}{c^4} \frac{M^2}{R} \frac{1}{d} e^{-\frac{t}{\tau}}~,
\end{align} 
where we explicitly put back the time dependence using equation~(\ref{alpha_decay}). 

One important use of equation~(\ref{GW_signal}) is that it can provide an upper limit on $\alpha_{2, 2}(0)$ upon the \emph{non-observation} of a GW signal (which produces an upper limit on $h_0(0)$). The corresponding value of $\alpha_{2, 2}(0)$ for a given value of $h_0(0)$ is
\begin{align}
\label{upper_limit_alpha}
\alpha_{2, 2}(0) \approx 1.4 \times 10^{-4} \left(\frac{M}{1.4~\text{M}_\odot}\right)^{-2} \left(\frac{R}{10~\text{km}}\right) \left(\frac{d}{1~\text{kpc}}\right) \left(\frac{h_0(0)}{1 \times 10^{-21}}\right)~,
\end{align} 
where we have used representative values as an example. An upper limit on $\alpha_{2, 2}(0)$ can be reported in burst searches, analogous to how upper limits are reported for $r$-modes \citep{owen2010, fesikPapa2020, LVK2021rmodes}. The constraint provided by upper limits on $\alpha_{2, 2}(0)$ could help falsify mode excitation models.

\subsection{Signal-to-noise ratio}

We will now take the GW signal from equation~(\ref{GW_signal}) and use it to calculate the SNR. Note that the GW signal provided is closed-form so unlike most burst searches, a fully modelled matched filter search is possible.

In general, one needs to have a large enough SNR to claim a GW detection, with the threshold value determined by how many false alarms and false dismissals one allows \citep[e.g.][]{jaranowskiKrolakSchutz1998, LVC2004threshold}. Once decided, the next consideration that affects the threshold is the type and width of the search. The wider the search, the higher the SNR threshold must be. To determine the exact threshold is a challenge in itself, e.g.~\citet{tenorioetal2022}, so we adopt the pragmatic approach of looking at the threshold calculated for an actual (published) GW search.  For this purpose, we choose the analysis of \citet{abadieetal2011}, who looked for modelled $f$-mode signals following a glitch in the Vela pulsar, which is essentially the sort of search relevant here.

As we will see shortly in equation~(\ref{SNR_decaying_exponential}), the SNR for an exponentially decaying sinusoidal signal depends on the initial GW amplitude, the decay time-scale of the signal and the detector sensitivity at GW frequency $f$. For their LIGO Science Run 5 (S5) data, \citet{abadieetal2011} found that at least 90\% of injected signals were detectable if $h_0(0) > 8.3 \times 10^{-21}$. The average injected decay time-scale was $\tau = 0.275~\text{s}$ and the average GW frequency was $f = 2~\text{kHz}$, which had a corresponding detector sensitivity of $\sqrt{S_n(f)} \sim 3 \times 10^{-22}~\text{Hz}^{-\frac{1}{2}}$ for S5 \citep{LSC2009}. Putting these numbers into equation~(\ref{SNR_decaying_exponential}) gives a SNR threshold of $\sim 10$, which we use as our nominal threshold for detection. 

In fact, the frequency and decay time-scale priors used in \citet{abadieetal2011} were relatively wide, and more recent observations from GW170817 have put stronger constraints on these quantities \citep{wenetal2019}\footnote{It should be noted that the mode damping time-scale of our model is consistent with the allowed time-scale range of \citet{wenetal2019} only when $R \gtrsim 12~\text{km}$. This is expected since real NSs, which the \cite{wenetal2019} analysis is based on, are thought to have radii closer to 12~km rather than the 10~km used in our canonical description.}. Using narrower priors will have the effect of decreasing the SNR threshold required given all else is the same. 
	
However, we have not yet considered the uncertainty in the start time of the GCs/AGCs. A greater error in the start time will increase the threshold needed for a detection. \citet{espinozaetal2014, espinozaetal2021} could only get the errors in the start time down to around a day, as temporal resolution was not their primary focus given they had decades of data to process (Antonopoulou \& Espinoza, private communication). This makes searching for the associated burst GWs more difficult, but not impossible. There are reasons to be optimistic though. Following \citet{espinozaetal2014, espinozaetal2021}, a re-analysis optimised for temporal resolution would provide more accurate and precise start times, potentially located to within a few hours. Furthermore, with recent glitch observations having resolutions of a few seconds to a few minutes \citep{palfreymanetal2018, shawetal2018}, it is not implausible to imagine future GCs and AGCs being observed with a similar resolution, especially when future detectors like the Square Kilometre Array will have dedicated programs for pulsar timing \citep{bailesetal2016}. In this ideal situation, the search for GWs from GCs/AGCs would be over a similar time window as used for glitches in \cite{abadieetal2011}, which had an ``on-source'' time window of $120~\text{s}$ corresponding to a window wide enough to contain errors of up to $3\sigma$ in the start time.

We will now find an analytical expression for the SNR expected from our model. We begin by following \citet{jaranowskiKrolakSchutz1998} in defining the (square of the) optimal SNR as
\begin{align}
\rho_0^2 \equiv 4 \int_{0}^{\infty}  \frac{|\tilde{h}(f)|^2}{S_\text{n}(f)} df ~,
\end{align}
where the integral is over all positive GW frequencies $f$, $S_\text{n}(f)$ is the (one-sided) power spectral density, $|\tilde{h}(f)|^2 = \tilde{h}^*(f)\tilde{h}(f)$ and the tilde represents a Fourier transform. The SNR is ``optimal'' as we are assuming the matched filter perfectly describes the GW signal. 

For a source that emits a GW signal whose amplitude changes but frequency remains fixed, such as a decaying oscillation mode, $S_\text{n}(f)$ remains constant so can be taken out of the integral. This leads to
\begin{align}
\label{SNR_FT}
\rho_0^2 = \frac{4}{S_\text{n}(f)} \int_{0}^{\infty}  |\tilde{h}(f)|^2 df~.
\end{align}
Note that the frequency in $S_\text{n}(f)$ is the GW frequency but the $f$ in the integral is only a dummy variable. Then, exploiting the fact that the integrand is an even function of $f$, and using Parseval's theorem\footnote{One can also explicitly calculate the Fourier transform of $h(t)$ to use in equation~(\ref{SNR_FT}) but using Parseval's theorem shortens the calculation. The final expression for the SNR is the same in both cases, as expected.}, we find
\begin{align}
\rho_0^2 = \frac{2}{S_\text{n}(f)} \int_{-\infty}^{\infty}  |h(t)|^2 dt~.
\end{align}
Using an exponentially decaying sinusoid (equation~(\ref{exponentially_decaying_sinusoid})) for $h(t)$, we find that the optimal SNR is
\begin{align}
\label{SNR_decaying_exponential}
\rho_0 = \frac{h_0(0) \sqrt{\tau}}{\sqrt{2S_\text{n}(f)}}~.
\end{align}
Putting in our specific solutions for $h_0(0)$ and $\tau$ from equations~(\ref{GW_signal}) and (\ref{tau_phys}) respectively, we find the SNR (squared) obtainable from our model, in terms of $\alpha_{2,2}(0)$, is
\begin{align}
\rho_0^2 = \frac{15}{2\pi} \alpha_{2, 2}^2(0) \frac{G}{c^3} \frac{1}{S_\text{n}(f)} M R^2 \frac{1}{d^2}~.
\end{align}
For a NS with known/estimated $M$, $R$ and $d$ and for a GW detector with power spectral density $S_\text{n}(f)$, the SNR becomes solely a function of $\alpha_{2,2}(0)$, i.e. depends only on how large the initial mode amplitude is. For representative values, we find
\begin{align}
\rho_0 \approx 1.3 \left(\frac{\alpha_{2, 2}(0) }{1 \times 10^{-6}} \right)  &\left(\frac{\sqrt{S_\text{n}(f)}}{1 \times 10^{-24}~\text{Hz}^{-\frac{1}{2}}} \right)^{-1} \left(\frac{M}{1.4~\text{M}_\odot} \right)^{\frac{1}{2}} \cdots \nonumber \\ 
&~~~~\cdots~\left(\frac{R}{10~\text{km}} \right) \left(\frac{d}{1~\text{kpc}} \right)^{-1}~,
\end{align}
where $\sqrt{S_\text{n}(f)} = 1 \times 10^{-24}~\text{Hz}^{-\frac{1}{2}}$ is the value at $f \approx 2~\text{kHz}$ for the Einstein Telescope (ET) \citep{hildetal2011}. The sensitivity of Cosmic Explorer 1 (CE) is essentially the same at this frequency, so we expect to obtain a similar SNR \citep{reitzeetal2019}.

Since we have a prescription of how to go from the mode amplitude to a change in spin frequency (equation~(\ref{glitch_size_alpha})), we can use that to directly find what the expected SNR (squared) would be if we observe a change in spin frequency
\begin{align}
\label{SNR_h_0}
\rho_0^2 = \frac{2\pi\sqrt{5G}}{c^3} \frac{1}{S_\text{n}(f)} M^{\frac{1}{2}} R^{\frac{7}{2}} d^{-2} \Delta \nu~.
\end{align}
Again, for representative values, we find
\begin{align}
\rho_0 \approx 2.7 &\left(\frac{\sqrt{S_\text{n}(f)}}{1 \times 10^{-24}~\text{Hz}^{-\frac{1}{2}}} \right)^{-1} \left(\frac{M}{1.4~\text{M}_\odot} \right)^{\frac{1}{4}} \cdots \nonumber \\ 
&~~~~\cdots~\left(\frac{R}{10~\text{km}} \right)^{\frac{7}{4}} \left(\frac{d}{1~\text{kpc}} \right)^{-1} \left(\frac{\Delta \nu}{1 \times 10^{-8}~\text{Hz}} \right)^{\frac{1}{2}}~.
\end{align}

\section{Applying the model to data}
\label{section_applying_to_data}

As mentioned previously, we are mainly interested in whether GCs and AGCs can be explained by our model. We have shown that changes in spin frequency map directly to an initial mode amplitude, with an $m>0$ mode being excited for a GC and an $m<0$ mode being excited for an AGC. We will calculate what these initial mode amplitudes are using actual electromagnetic data in Section~\ref{subsection_initial_mode_amplitude}. Then, in Section~\ref{subsection_signal_to_noise_ratio}, we will assume the model presented here really is the mechanism behind GCs and AGCs and explore whether the GWs given off will be detectable or not.

The data we have are values of $\Delta \nu$ for the GCs and AGCs of the Crab and Vela pulsars, which were kindly provided by C.~M.~Espinoza \citep{espinozaetal2014, espinozaetal2021}. Even though we are mainly focusing on GCs and AGCs, we will speculate and extend the model to include glitches too, as we will soon find out that current upper limits on $h_0(0)$ allow us to do so. Values of $\Delta \nu$ for glitches have been taken from the JBCA Glitch Catalogue \citep{espinozaetal2011}. Additionally, we are required to know the distances to the Crab and Vela pulsars, which are taken from the ATNF Pulsar Catalogue \citep{manchesteretal2005}, and are $d = 2.00~\text{kpc}$ and $d = 0.28~\text{kpc}$ respectively. We will assume $m = \pm 2$ here as we assume the excitation of such non-axisymmetric quadrupolar modes, as was described in Section \ref{section_the_model}.  We will also always use the canonical values of $M = 1.4~\text{M}_\odot$ and $R = 10~\text{km}$.

\subsection{Initial mode amplitude}
\label{subsection_initial_mode_amplitude}

\begin{figure}
	\centering
	\includegraphics[width=\linewidth]{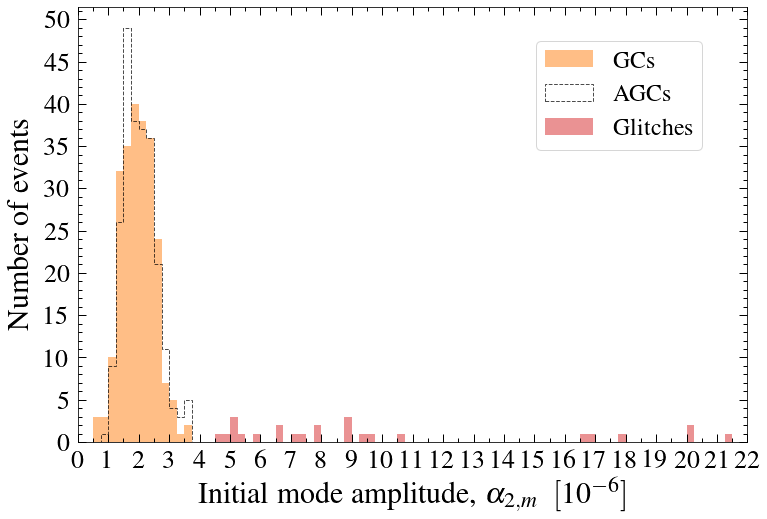}
	\caption[Histogram of mode amplitudes for the Crab's GCs, AGCs and glitches]{
		\label{fig_alpha_crab}
		Histogram showing initial mode excitation amplitudes required to explain GCs, AGCs and glitches of the Crab pulsar. Some values are not shown for clarity. These are at values of $\alpha_{2, m} = (32.2, 52.3, 81.2) \times 10^{-6}$, all of which belongs to glitches.
	}
\end{figure}

\begin{figure}
	\centering
	\includegraphics[width=\linewidth]{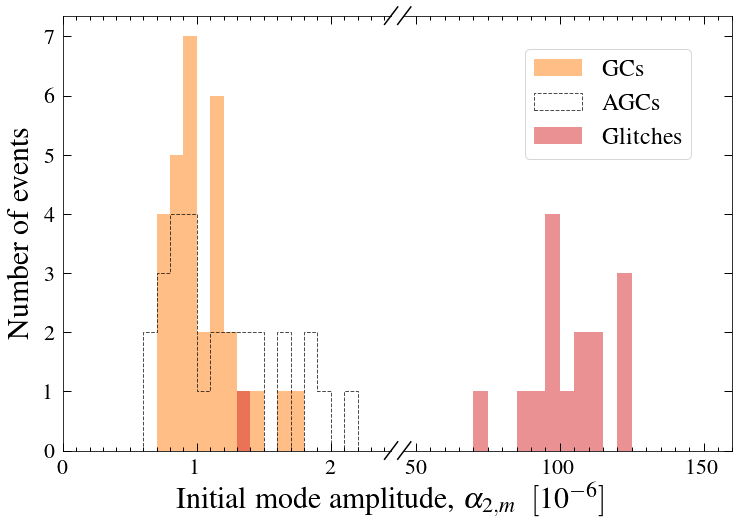}
	\caption[Histogram of mode amplitudes for Vela's GCs, AGCs and glitches]{
		\label{fig_alpha_vela}
		Histogram showing initial mode excitation amplitudes required to explain GCs, AGCs and glitches of the Vela pulsar. Note the change in scale after the break on the $x$-axis. Some values are not shown for clarity. These are at values of $\alpha_{2, m} = (7.6, 31.0) \times 10^{-6}$, both of which belongs to glitches.
	}
\end{figure}

We first use equation~(\ref{glitch_size_alpha}) to calculate the initial mode excitation amplitudes required to explain the small changes in spin observed in GCs ($m = +2$) and AGCs ($m=-2$). This is shown as histograms in Figure~\ref{fig_alpha_crab} for the Crab pulsar and Figure~\ref{fig_alpha_vela} for the Vela pulsar. We see that both GCs and AGCs require a similar initial mode amplitude, with the Crab requiring $\alpha_{2, m}\approx 2 \times 10^{-6}$ and Vela requiring $\alpha_{2, m}\approx 1 \times 10^{-6}$. This corresponds to a mode amplitude that is one millionth of the radius of the NS, which is about $1~\text{cm}$.

Although the theory behind glitches is fairly well developed \cite[e.g.][]{ruderman1969, andersonItoh1975, haskellMelatos2015}, we will apply the calculations to glitches too.  In other words, we will explore the idea that glitches represents the excitation and decay of a relatively larger mode, that propagates in the negative mathematical sense. We regard this as extremely speculative, given the existing perfectly plausible models for pulsar glitches \citep{l_g-s_12}.
 The required initial mode excitation amplitudes that our model then gives for the Crab and Vela's glitches are shown in Figures~\ref{fig_alpha_crab} and \ref{fig_alpha_vela} respectively. They have values of $\alpha_{2, 2} \sim 1 \times 10^{-5}$ for the Crab and $\alpha_{2, 2} \lesssim 1.3 \times 10^{-4}$ for Vela.

\subsection{Signal-to-noise ratio}
\label{subsection_signal_to_noise_ratio}

\begin{figure}
	\centering
	\includegraphics[width=\linewidth]{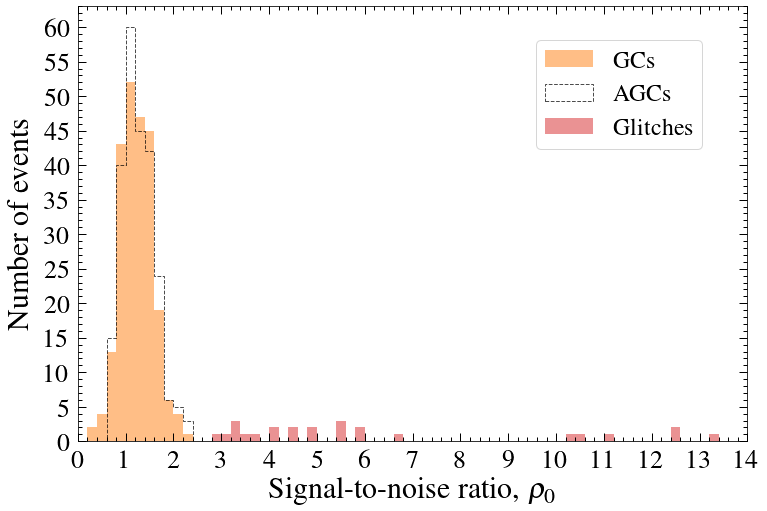}
	\caption[Histogram of SNRs for the Crab's GCs, AGCs and glitches for the ET]{
		\label{fig_SNR_crab_ET}
		Histogram showing the SNR attainable by the ET (or CE) for the predicted GWs from GCs, AGCs and glitches for the Crab pulsar. Some SNRs are not shown for clarity. These are at values of: 20.0, 32.6 and 50.6, all of which belongs to glitches.
	}
\end{figure}

\begin{figure}
	\centering
	\includegraphics[width=\linewidth]{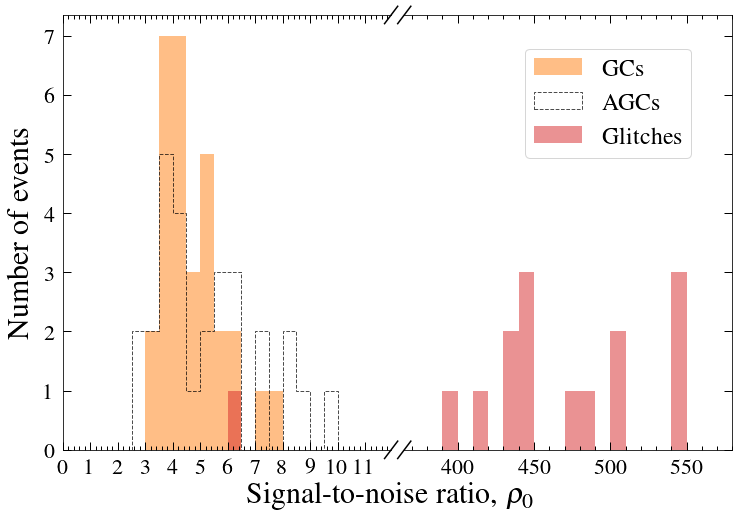}
	\caption[Histogram of SNRs for Vela's GCs, AGCs and glitches for the ET]{
		\label{fig_SNR_vela_ET}
		Histogram shows the SNR attainable by the ET (or CE) for the predicted GWs from GCs, AGCs and glitches for the Vela pulsar. Note the change in scale after the break on the $x$-axis. Some SNRs are not shown for clarity. These are at values of: 33.9, 138.0 and 330.9, all of which belongs to glitches.
	}
\end{figure}

This section shows the results of the GW SNR calculation when our model is applied. The relevant equation is equation~(\ref{SNR_h_0}) which takes the change in spin frequency as the primary input. The results of this calculation are shown in Figures~\ref{fig_SNR_crab_ET} and \ref{fig_SNR_vela_ET} which are histograms of the SNR for the Crab and Vela, respectively, assuming ET sensitivity \citep{hildetal2011}. The sensitivity of CE is very similar to the ET at GW frequencies of $f\approx 2~\text{kHz}$, so we can take the SNR values in the figures as representative of CE too \citep{reitzeetal2019}.

For the Crab, GCs and AGCs have a SNR $\sim 1$, whereas the SNR is around $\sim 5$ for Vela. It appears that GWs from individual GCs or AGCs will not be detectable with the ET or CE. However, combining the signal from both the ET and CE could improve the SNR by a factor of $\sqrt{2}$ since the two detectors are independent. Moreover, it might be possible to coherently stack multiple burst signals which improves the SNR by a factor of $\sqrt{N_\text{excite}}$, with $N_\text{excite}$ being the number of mode excitation events. This means that, for GCs/AGCs from Vela being detected by the network of ET and CE, one would need to stack 2+ events before the combined signal has a SNR that exceeds our nominal detection threshold, which is 10, but this depends greatly on the type and width of the search \cite[e.g.][]{walshetal2016}.

\begin{figure}
	\centering
	\includegraphics[width=\linewidth]{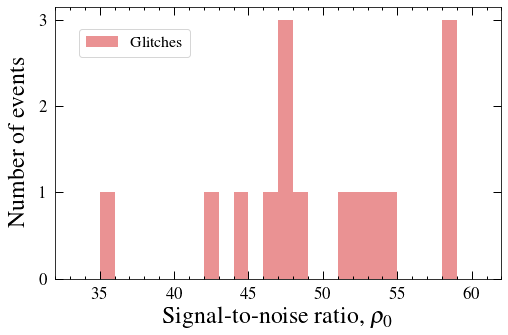}
	\caption[Histogram of SNRs for Vela's GCs, AGCs and glitches for aLIGO]{
		\label{fig_SNR_vela_aLIGO}
		Histogram showing the SNR attainable by Advanced LIGO at design sensitivity for the predicted GWs from glitches for the Vela pulsar. Some SNRs are not shown for clarity. These are at values of: 0.7, 3.7 and 14.9.
	}
\end{figure}

Moving onto the speculative case of applying our model to glitches, the suggestion is that glitches are caused by the excitation and decay of an $m = 2$ mode instead of vortex unpinning or starquakes. The SNRs for glitches (using the ET or CE) are also given in Figures~\ref{fig_SNR_crab_ET} and \ref{fig_SNR_vela_ET}. Some of the Crab's largest glitches and all of Vela's glitches should be detectable with the ET or CE if this suggestion is to be believed. In fact, Vela's glitch SNRs for the ET (or CE) are so large that we can consider what they would be for Advanced LIGO \citep[at design sensitivity,][]{aasietal2015}. The results of this are shown in Figure~\ref{fig_SNR_vela_aLIGO}.

If a signal is not detected, one might ask what we could learn from this. As mentioned in Section~\ref{subsection_gravitational_wave_strain}, an upper limit on $h_0(0)$ places an upper limit on $\alpha_{2, 2}(0)$. The most recent relevant (i.e. considers the same time-scales) study comes from a burst search conducted on Vela's August 2006 glitch which yielded an upper limit of $h_0(0) < 6.3 \times 10^{-21}$ \citep{abadieetal2011}. Using equation~(\ref{upper_limit_alpha}) and $d = 0.28~\text{kpc}$ \citep{manchesteretal2005}, we find this corresponds to an upper limit of $\alpha_{2, 2}(0) < 2.6 \times 10^{-4}$.

Note that the 2006 glitch itself had an observed magnitude of $\frac{\Delta \Omega}{\Omega} = 2.6 \times 10^{-6}$.  Using equation~(\ref{glitch_size_alpha}) this corresponds to an initial excitation amplitude of $\alpha_{2, 2}(0) = 1.1 \times 10^{-4}$, and using equation~(\ref{upper_limit_alpha}), to  a GW amplitude $h_0(0) = 2.7 \times 10^{-21}$.  We therefore see that the direct upper limit on GW emission following the 2006 glitch was not constraining for our model, but only by a factor of $2$ or so.  The Advanced LIGO detector sensitivity will be significantly better than that of S5, consistent with the detectable SNRs reported in Figure \ref{fig_SNR_vela_aLIGO}. 


\section{Energetics}
\label{section_energetics}

We will now consider the energetics of the model. We will consider the energy budget required to excite the GCs and AGCs at the observed rate and at the amplitudes calculated within our model.  Then, as a simple example, we will see whether this power can be provided from elasticity during the usual secular spin-down of a NS. Both quantities will be given as a fraction of the spin-down power, $\dot{E}_\text{spin-down} = - 4 \pi^2 I \nu \dot{\nu}$.

In the following subsections, it will be useful to know how many GCs/AGCs there were and over what timespan they occurred. For the Crab, there were 381 GCs and 383 AGCs (for a total of 764 events), and they occurred over a timespan of $T_\text{obs} \approx 10620~\text{d}$ \citep{espinozaetal2014}. Likewise, for Vela, there were 83 GCs and 66 AGCs (for a total of 149 events) that occurred over a timespan of $T_\text{obs} = 6865~\text{d}$ \citep{espinozaetal2021}. Also, so that everything is in one place, we have for the Crab: $\nu = 29.6~\text{Hz}$ and $\dot{\nu} = - 3.68\times10^{-10}~\text{Hz s}^{-1}$, and for Vela: $\nu = 11.2~\text{Hz}$ and $\dot{\nu} = - 1.56\times10^{-11}~\text{Hz s}^{-1}$ \citep{manchesteretal2005}.

\subsection{Power required for modes}
\label{subsection_power_required}

The power required to excite the modes, averaged over times long compared to the interval between events,  is simply given by the average mode energy, $\langle \delta E \rangle$, times the average frequency at which the modes are excited, $\mathcal{F} = N / T_\text{obs}$, where $N$ is the total number of GCs and AGCs observed across time $T_\text{obs}$. Explicitly, this is
\begin{align}
\langle \dot{E}_\text{mode} \rangle = \mathcal{F} \langle \delta E \rangle ~.
\end{align}
In \cite{yimJones2022}, we found an analytic expression for the mode energy which came from summing together kinetic, gravitational and internal energy contributions, and this gave
\begin{align}
\delta E = \alpha_{2,2}^2 \bar\rho \omega_2^2 R^5 = \frac{3}{5\pi} \alpha^2_{2,2} \frac{G M^2} {R}~,
\end{align}
meaning the average power required is
\begin{align}
\label{mode_power}
\langle \dot{E}_\text{mode} \rangle = \frac{3}{5\pi} \langle \alpha^2_{2,2} \rangle \frac{G M^2} {R} \mathcal{F}~. 
\end{align}
Putting in representative values, we find
\begin{align}
\langle \dot{E}_\text{mode} \rangle \approx 3.9 \times 10^{34} \left(\frac{\sqrt{\langle \alpha^2_{2,2} \rangle}}{1 \times 10^{-6}} \right)^2 &\left(\frac{M}{1.4~\text{M}_\odot} \right)^2 \left(\frac{R}{10~\text{km}} \right)^{-1} \cdots \nonumber \\ 
& \cdots \left(\frac{\mathcal{F}}{1/(30~\text{d})} \right) ~ \text{erg~s}^{-1}~.
\end{align}
Note that this is the time-averaged power required to excite the \emph{observed} GC/AGC events. The detection of these events in radio pulsar data partially depends on observational cadence, and is currently limited by timing accuracy \citep{espinozaetal2014, espinozaetal2021}. For telescopes with better timing accuracy, one might be able to observe a greater number of small events in the same amount of time, leading to a larger value for $\langle \dot{E}_\text{mode} \rangle$, correspondingly 
increasing the required energy budget.

We will now look at the specific cases of the Crab and Vela. Taking the $\Delta \nu$ data for GCs and AGCs and inputting into equation~(\ref{glitch_size_alpha}), we find that $\sqrt{\langle \alpha^2_{2,2} \rangle} = 2.1 \times 10^{-6}$ for the Crab and $\sqrt{\langle \alpha^2_{2,2} \rangle} = 1.2 \times 10^{-6}$ for Vela. After calculating event frequencies $\mathcal{F}$, we find that the Crab requires an average mode power of $\langle \dot{E}_\text{mode} \rangle \sim 7 \times 10^{-4}~\dot{E}_\text{spin-down}$ and Vela requires $\langle \dot{E}_\text{mode} \rangle \sim 4 \times 10^{-3}~\dot{E}_\text{spin-down}$. In other words, we need less than 1\% of the spin-down power to sustain the excitation of modes as frequently as they appear in observations.  Given that all mode energy is radiated away as GWs, this means that about 0.07\% - 0.4\% of the spin-down power goes into GW emission if this model is to be believed. 

\subsection{Power from elasticity}

The question now is where can the power calculated in the previous subsection come from? A natural mechanism to look at is the build up of elastic energy that is stored in the crust as the NS spins down. Here, we will do a rough calculation to see how much power can be extracted from the elasticity of the crust, assuming the crust becomes maximally strained.

One can imagine that at some point in time, a NS is rotating at some angular velocity which gives it an oblateness. As the NS spins down, the oblateness wants to decrease due to a weakening centrifugal force, but a solid crust prevents it from doing so fully. This then strains the crust so we get a build up of elastic energy that we could harness for the excitation of our modes. The explanation provided here forms part of the starquake model \citep{baymPines1971}.

Therefore, we will use the same simple equations provided by \cite{baymPines1971} to help us with our calculation. The elastic energy can be written as
\begin{align}
E_\text{el} = B(\varepsilon_\text{ref} - \varepsilon)^2~,
\end{align}
where $B$ is a constant that depends on the shear modulus of the crust \citep[e.g.][]{ogataIchimaru1990, strohmayeretal1991, keerJones2015}, and the oblateness can be written as
\begin{align}
\varepsilon = \frac{I_\text{sph} \Omega^2}{4(A + B)} + \frac{B}{A + B}\varepsilon_{\text{ref}}~,
\end{align}
where $A$ is another constant that parametrises how the gravitational potential energy changes as the oblateness varies. For an incompressible canonical NS, $B/A \sim 10^{-5}$ \citep{baymPines1971}. $\varepsilon_\text{ref}$ is the reference oblateness which is the oblateness when there is no strain in the crust. For the situation described above, this means $\varepsilon_\text{ref} > \varepsilon$. One can then differentiate the two equations above with respect to time to find
\begin{align}
\label{rate_elastic_energy}
\dot{E}_\text{el} = - 2 B \dot{\varepsilon} (\varepsilon_\text{ref} - \varepsilon)
\end{align}
and
\begin{align}
\dot{\varepsilon} = \frac{I_\text{sph} \Omega \dot{\Omega}}{2(A + B)} = \frac{\dot{E}_\text{spin-down}}{2(A+B)}
\end{align}
where we have assumed the reference oblateness does not change between excitation events, i.e. $\dot{\varepsilon}_\text{ref} = 0$ (no plastic flow). We then substitute $\dot{\varepsilon}$ into equation~(\ref{rate_elastic_energy}) and use the fact that the largest strain the NS is able to endure is the breaking strain, $\varepsilon_\text{ref} - \varepsilon = u_\text{break}$, which has a value no larger than 0.1 \citep{horowitzKadau2009, baikoChugunov2018}. This means the maximum power we can get from elasticity is
\begin{align}
\frac{|\dot{E}_\text{el}|}{\dot{E}_\text{spin-down}} = \frac{B}{A+B} u_\text{break} \sim 10^{-6}~.
\end{align}
Clearly this is around 3 orders of magnitude too small to power the modes and so elasticity alone, as we have modelled it, cannot be the driver of the modes. In a more realistic situation, one might expect plastic flow to occur  \citep[e.g.][]{baikoChugunov2018}, perhaps preventing strains as large as $\sim 0.1$ being attained, further limiting the elastic energy available. 

On the other hand, one would also expect the NS to contain a superfluid, so like with glitches, some energy may be harnessed from there. Indeed, as is well known, the superfluid pinning model allows for much more energetic events than the starquake model \citep[e.g.][]{l_g-s_12}.  We will explore this idea in the future. Ultimately, we only require less than 1\% of the spin-down power to go into exciting the modes in the model, as shown in Section~\ref{subsection_power_required}. 

\section{Summary and Discussion}
\label{section_summary}

In this paper, we proposed a novel model for explaining the recently observed small spin-ups and spin-downs of the Crab and Vela pulsars, also known as glitch candidates and anti-glitch candidates \citep{espinozaetal2014, espinozaetal2021}. In the proposed model, we ascribe the change in spin frequency to the excitation and decay of a non-axisymmetric $f$-mode which propagates either against rotation (for a spin-up) or with rotation (for a spin-down).  For a given NS, the amplitude of  mode excitation is the only free parameter in the model, and can be calculated from  the observed change in spin frequency.

One of the key unique features of the model is the connection to GWs. The propagation of non-axisymmetric modes causes a time-varying mass quadrupole moment, and this generates GWs. We calculated details of the GW emission and its back-reaction on the NS in our previous work \citep{yimJones2022}.  We applied those results here to the excitation of oscillations for a spinning down NS. We used the predicted GW signal to assess whether GWs from this model were detectable or not, and gave expressions for the SNR as a function of the change in spin frequency. In particular, for a nominal SNR threshold of 10, signals from the Vela pulsar may be detectable by the ET or CE, but only by coherently combining several such signals. 

There are many improvements that can be made to the model, but the aim of this paper was to provide simple analytic results and a proof of concept, based on the most essential ingredients. Nevertheless, future studies should focus on removing some of the simplifying assumptions. This includes: modelling exactly how the modes are excited, which is closely related to the energy budget problem, calculating the next leading order corrections due to rotation, modelling a realistic interior with a specified equation of state, and, looking at the effect of higher order modes. 

Besides the GWs aspect, one might wonder how NS oscillations may influence other electromagnetic observations like individual pulse profiles. Taking the Crab as an example; it has a period of around $30~\text{ms}$ and a pulse fraction of around 3\% \citep{gouldLyne1998}, meaning each pulse lasts around $1~\text{ms}$ or around 2 $f$-mode periods. The presence of $f$-modes could therefore ``shake'' the magnetosphere and superimpose substructure to the pulse profiles of individual pulses, until the $f$-mode decays away ($\sim$few rotations). This would be observed as a handful of individual pulses suddenly developing ``subpulses'' which would then recover quickly back to normal. 

Furthermore, the sudden increase or decrease to the angular frequency of the NS has implications for the coupling to other parts of the NS, particularly when we consider the $\Delta \dot{\nu}$ observed for GCs and AGCs. We will comment more on this in a separate publication.

Finally, there is a natural extension which first inspired us to tackle this problem, and it is whether the excitation and decay of $f$-modes could explain long-term timing noise. The idea is that there are consecutive, and perhaps unresolvable, spin-ups and spin-downs which collectively give rise to timing noise. This is somewhat similar to early theories of how microglitches cause timing noise \citep[e.g.][]{cheng1987, chengetal1988}, but with the model presented here, it would be testable with GW observations. If timing noise is characterised by changes in spin frequency only, then the model presented here is sufficient, but there needs to be more thought put into how the modes are consecutively excited and why a mode propagates in a certain direction as opposed to the opposite direction. We will leave these unanswered questions open for now as they will form the basis of future studies.

\section*{Acknowledgements}

The authors would like to thank Danai Antonopoulou and Crist\'{o}bal Espinoza for helpful discussions regarding glitch candidates and anti-glitch candidates and for providing us with their data. We would also like to thank David Keitel, Luana Modafferi, Joan Moragues and Rodrigo Tenorio for their guidance on detectability thresholds. Finally, we are grateful for the detailed discussions with Ian Hawke and Bernard Schutz which helped with the clarity of this work. GY acknowledges support from the EPSRC via grant number EP/N509747/1. DIJ acknowledges support from the STFC via grant number ST/R00045X/1. This paper has been assigned document number LIGO-P2200096.

\section*{Data Availability}

The data used in this article was cited accordingly in the main text. The main sources used were the JBCA Glitch Catalogue \citep{espinozaetal2011} and the ATNF Pulsar Catalogue \citep{manchesteretal2005}. Glitch candidate and anti-glitch candidate data were provided by C.~M.~Espinoza.



\bibliographystyle{mnras}
\bibliography{references} 

\begin{thebibliography}{}
\makeatletter
\relax
\def\mn@urlcharsother{\let\do\@makeother \do\$\do\&\do\#\do\^\do\_\do\%\do\~}
\def\mn@doi{\begingroup\mn@urlcharsother \@ifnextchar [ {\mn@doi@}
  {\mn@doi@[]}}
\def\mn@doi@[#1]#2{\def\@tempa{#1}\ifx\@tempa\@empty \href
  {http://dx.doi.org/#2} {doi:#2}\else \href {http://dx.doi.org/#2} {#1}\fi
  \endgroup}
\def\mn@eprint#1#2{\mn@eprint@#1:#2::\@nil}
\def\mn@eprint@arXiv#1{\href {http://arxiv.org/abs/#1} {{\tt arXiv:#1}}}
\def\mn@eprint@dblp#1{\href {http://dblp.uni-trier.de/rec/bibtex/#1.xml}
  {dblp:#1}}
\def\mn@eprint@#1:#2:#3:#4\@nil{\def\@tempa {#1}\def\@tempb {#2}\def\@tempc
  {#3}\ifx \@tempc \@empty \let \@tempc \@tempb \let \@tempb \@tempa \fi \ifx
  \@tempb \@empty \def\@tempb {arXiv}\fi \@ifundefined
  {mn@eprint@\@tempb}{\@tempb:\@tempc}{\expandafter \expandafter \csname
  mn@eprint@\@tempb\endcsname \expandafter{\@tempc}}}

\bibitem[\protect\citeauthoryear{{Aasi} et~al.,}{{Aasi}
  et~al.}{2015}]{aasietal2015}
{Aasi} J.,  et~al., 2015, \mn@doi [Classical and Quantum Gravity]
  {10.1088/0264-9381/32/7/074001}, \href
  {https://ui.adsabs.harvard.edu/abs/2015CQGra..32g4001L} {32, 074001}

\bibitem[\protect\citeauthoryear{{Abadie} et~al.,}{{Abadie}
  et~al.}{2011}]{abadieetal2011}
{Abadie} J.,  et~al., 2011, \mn@doi [\prd] {10.1103/PhysRevD.83.042001}, \href
  {https://ui.adsabs.harvard.edu/abs/2011PhRvD..83d2001A} {83, 042001}

\bibitem[\protect\citeauthoryear{{Abbott} et~al.,}{{Abbott}
  et~al.}{2004}]{LVC2004threshold}
{Abbott} B.,  et~al., 2004, \mn@doi [\prd] {10.1103/PhysRevD.69.082004}, \href
  {https://ui.adsabs.harvard.edu/abs/2004PhRvD..69h2004A} {69, 082004}

\bibitem[\protect\citeauthoryear{{Abbott} et~al.,}{{Abbott}
  et~al.}{2009}]{LSC2009}
{Abbott} B.~P.,  et~al., 2009, \mn@doi [\prd] {10.1103/PhysRevD.80.102002},
  \href {https://ui.adsabs.harvard.edu/abs/2009PhRvD..80j2002A} {80, 102002}

\bibitem[\protect\citeauthoryear{{Abbott} et~al.,}{{Abbott}
  et~al.}{2016}]{LVC2016firstdetection}
{Abbott} B.~P.,  et~al., 2016, \mn@doi [\prl] {10.1103/PhysRevLett.116.061102},
  \href {https://ui.adsabs.harvard.edu/abs/2016PhRvL.116f1102A} {116, 061102}

\bibitem[\protect\citeauthoryear{{Abbott} et~al.,}{{Abbott}
  et~al.}{2020}]{LVK2020prospectsreview}
{Abbott} B.~P.,  et~al., 2020, \mn@doi [Living Reviews in Relativity]
  {10.1007/s41114-020-00026-9}, \href
  {https://ui.adsabs.harvard.edu/abs/2020LRR....23....3A} {23, 3}

\bibitem[\protect\citeauthoryear{{Abbott} et~al.,}{{Abbott}
  et~al.}{2021a}]{LVK2021allskyshortburstsearch}
{Abbott} R.,  et~al., 2021a, \mn@doi [\prd] {10.1103/PhysRevD.104.122004},
  \href {https://ui.adsabs.harvard.edu/abs/2021PhRvD.104l2004A} {104, 122004}

\bibitem[\protect\citeauthoryear{{Abbott} et~al.,}{{Abbott}
  et~al.}{2021b}]{LVK2021rmodes}
{Abbott} R.,  et~al., 2021b, \mn@doi [\apj] {10.3847/1538-4357/ac0d52}, \href
  {https://ui.adsabs.harvard.edu/abs/2021ApJ...922...71A} {922, 71}

\bibitem[\protect\citeauthoryear{{Alpar}, {Nandkumar}  \& {Pines}}{{Alpar}
  et~al.}{1986}]{alparNandkumarPines1986}
{Alpar} M.~A.,  {Nandkumar} R.,   {Pines} D.,  1986, \mn@doi [\apj]
  {10.1086/164765}, \href
  {https://ui.adsabs.harvard.edu/abs/1986ApJ...311..197A} {311, 197}

\bibitem[\protect\citeauthoryear{{Anderson} \& {Itoh}}{{Anderson} \&
  {Itoh}}{1975}]{andersonItoh1975}
{Anderson} P.~W.,  {Itoh} N.,  1975, \mn@doi [\nat] {10.1038/256025a0}, \href
  {http://adsabs.harvard.edu/abs/1975Natur.256...25A} {256, 25}

\bibitem[\protect\citeauthoryear{{Andersson}}{{Andersson}}{2021}]{andersson2021}
{Andersson} N.,  2021, \mn@doi [Universe] {10.3390/universe7040097}, \href
  {https://ui.adsabs.harvard.edu/abs/2021Univ....7...97A} {7, 97}

\bibitem[\protect\citeauthoryear{{Andersson} \& {Kokkotas}}{{Andersson} \&
  {Kokkotas}}{1998}]{anderssonKokkotas1998}
{Andersson} N.,  {Kokkotas} K.~D.,  1998, \mn@doi [\mnras]
  {10.1046/j.1365-8711.1998.01840.x}, \href
  {https://ui.adsabs.harvard.edu/abs/1998MNRAS.299.1059A} {299, 1059}

\bibitem[\protect\citeauthoryear{{Baiko} \& {Chugunov}}{{Baiko} \&
  {Chugunov}}{2018}]{baikoChugunov2018}
{Baiko} D.~A.,  {Chugunov} A.~I.,  2018, \mn@doi [\mnras]
  {10.1093/mnras/sty2259}, \href
  {https://ui.adsabs.harvard.edu/abs/2018MNRAS.480.5511B} {480, 5511}

\bibitem[\protect\citeauthoryear{{Bailes} et~al.,}{{Bailes}
  et~al.}{2016}]{bailesetal2016}
{Bailes} M.,  et~al., 2016, in MeerKAT Science: On the Pathway to the SKA.
  p.~11 (\mn@eprint {arXiv} {1803.07424})

\bibitem[\protect\citeauthoryear{{Basu} et~al.,}{{Basu}
  et~al.}{2022}]{basuetal2022}
{Basu} A.,  et~al., 2022, \mn@doi [\mnras] {10.1093/mnras/stab3336}, \href
  {https://ui.adsabs.harvard.edu/abs/2022MNRAS.510.4049B} {510, 4049}

\bibitem[\protect\citeauthoryear{{Baym} \& {Pines}}{{Baym} \&
  {Pines}}{1971}]{baymPines1971}
{Baym} G.,  {Pines} D.,  1971, \mn@doi [Annals of Physics]
  {10.1016/0003-4916(71)90084-4}, \href
  {https://ui.adsabs.harvard.edu/abs/1971AnPhy..66..816B} {66, 816}

\bibitem[\protect\citeauthoryear{{Cheng}}{{Cheng}}{1987}]{cheng1987}
{Cheng} K.~S.,  1987, \mn@doi [\apj] {10.1086/165673}, \href
  {https://ui.adsabs.harvard.edu/abs/1987ApJ...321..805C} {321, 805}

\bibitem[\protect\citeauthoryear{{Cheng}, {Alpar}, {Pines}  \&
  {Shaham}}{{Cheng} et~al.}{1988}]{chengetal1988}
{Cheng} K.~S.,  {Alpar} M.~A.,  {Pines} D.,   {Shaham} J.,  1988, \mn@doi
  [\apj] {10.1086/166517}, \href
  {https://ui.adsabs.harvard.edu/abs/1988ApJ...330..835C} {330, 835}

\bibitem[\protect\citeauthoryear{{Clark}, {Bauswein}, {Stergioulas}  \&
  {Shoemaker}}{{Clark} et~al.}{2016}]{clarketal2016}
{Clark} J.~A.,  {Bauswein} A.,  {Stergioulas} N.,   {Shoemaker} D.,  2016,
  \mn@doi [Classical and Quantum Gravity] {10.1088/0264-9381/33/8/085003},
  \href {https://ui.adsabs.harvard.edu/abs/2016CQGra..33h5003C} {33, 085003}

\bibitem[\protect\citeauthoryear{{Cordes} \& {Downs}}{{Cordes} \&
  {Downs}}{1985}]{cordesDowns1985}
{Cordes} J.~M.,  {Downs} G.~S.,  1985, \mn@doi [\apjs] {10.1086/191076}, \href
  {https://ui.adsabs.harvard.edu/abs/1985ApJS...59..343C} {59, 343}

\bibitem[\protect\citeauthoryear{{Cordes} \& {Greenstein}}{{Cordes} \&
  {Greenstein}}{1981}]{cordesGreenstein1981}
{Cordes} J.~M.,  {Greenstein} G.,  1981, \mn@doi [\apj] {10.1086/158883}, \href
  {https://ui.adsabs.harvard.edu/abs/1981ApJ...245.1060C} {245, 1060}

\bibitem[\protect\citeauthoryear{{Cordes}, {Downs}  \&
  {Krause-Polstorff}}{{Cordes} et~al.}{1988}]{cordesDownsKrause-Polstroff1988}
{Cordes} J.~M.,  {Downs} G.~S.,   {Krause-Polstorff} J.,  1988, \mn@doi [\apj]
  {10.1086/166518}, \href {http://adsabs.harvard.edu/abs/1988ApJ...330..847C}
  {330, 847}

\bibitem[\protect\citeauthoryear{{Corsi} \& {Owen}}{{Corsi} \&
  {Owen}}{2011}]{corsiOwen2011}
{Corsi} A.,  {Owen} B.~J.,  2011, \mn@doi [\prd] {10.1103/PhysRevD.83.104014},
  \href {https://ui.adsabs.harvard.edu/abs/2011PhRvD..83j4014C} {83, 104014}

\bibitem[\protect\citeauthoryear{{Espinoza}, {Lyne}, {Stappers}  \&
  {Kramer}}{{Espinoza} et~al.}{2011}]{espinozaetal2011}
{Espinoza} C.~M.,  {Lyne} A.~G.,  {Stappers} B.~W.,   {Kramer} M.,  2011,
  \mn@doi [\mnras] {10.1111/j.1365-2966.2011.18503.x}, \href
  {http://adsabs.harvard.edu/abs/2011MNRAS.414.1679E} {414, 1679}

\bibitem[\protect\citeauthoryear{{Espinoza}, {Antonopoulou}, {Stappers},
  {Watts}  \& {Lyne}}{{Espinoza} et~al.}{2014}]{espinozaetal2014}
{Espinoza} C.~M.,  {Antonopoulou} D.,  {Stappers} B.~W.,  {Watts} A.,   {Lyne}
  A.~G.,  2014, \mn@doi [\mnras] {10.1093/mnras/stu395}, \href
  {http://adsabs.harvard.edu/abs/2014MNRAS.440.2755E} {440, 2755}

\bibitem[\protect\citeauthoryear{{Espinoza}, {Antonopoulou}, {Dodson},
  {Stepanova}  \& {Scherer}}{{Espinoza} et~al.}{2021}]{espinozaetal2021}
{Espinoza} C.~M.,  {Antonopoulou} D.,  {Dodson} R.,  {Stepanova} M.,
  {Scherer} A.,  2021, \mn@doi [\aap] {10.1051/0004-6361/202039044}, \href
  {https://ui.adsabs.harvard.edu/abs/2021A&A...647A..25E} {647, A25}

\bibitem[\protect\citeauthoryear{{Ferrari}, {Miniutti}  \& {Pons}}{{Ferrari}
  et~al.}{2003}]{ferrariMiniuttiPons2003}
{Ferrari} V.,  {Miniutti} G.,   {Pons} J.~A.,  2003, \mn@doi [\mnras]
  {10.1046/j.1365-8711.2003.06580.x}, \href
  {https://ui.adsabs.harvard.edu/abs/2003MNRAS.342..629F} {342, 629}

\bibitem[\protect\citeauthoryear{{Fesik} \& {Papa}}{{Fesik} \&
  {Papa}}{2020}]{fesikPapa2020}
{Fesik} L.,  {Papa} M.~A.,  2020, \mn@doi [\apj] {10.3847/1538-4357/ab8193},
  \href {https://ui.adsabs.harvard.edu/abs/2020ApJ...895...11F} {895, 11}

\bibitem[\protect\citeauthoryear{{Flores}, {Hall}  \& {Jaikumar}}{{Flores}
  et~al.}{2017}]{floresHallJaikumar2017}
{Flores} C.~V.,  {Hall} Z.~B.,   {Jaikumar} P.,  2017, \mn@doi [\prc]
  {10.1103/PhysRevC.96.065803}, \href
  {https://ui.adsabs.harvard.edu/abs/2017PhRvC..96f5803F} {96, 065803}

\bibitem[\protect\citeauthoryear{{Gould} \& {Lyne}}{{Gould} \&
  {Lyne}}{1998}]{gouldLyne1998}
{Gould} D.~M.,  {Lyne} A.~G.,  1998, \mn@doi [\mnras]
  {10.1046/j.1365-8711.1998.02018.x}, \href
  {https://ui.adsabs.harvard.edu/abs/1998MNRAS.301..235G} {301, 235}

\bibitem[\protect\citeauthoryear{{Gualtieri}, {Kantor}, {Gusakov}  \&
  {Chugunov}}{{Gualtieri} et~al.}{2014}]{gualtierietal2014}
{Gualtieri} L.,  {Kantor} E.~M.,  {Gusakov} M.~E.,   {Chugunov} A.~I.,  2014,
  \mn@doi [\prd] {10.1103/PhysRevD.90.024010}, \href
  {https://ui.adsabs.harvard.edu/abs/2014PhRvD..90b4010G} {90, 024010}

\bibitem[\protect\citeauthoryear{{Haensel}, {Zdunik}, {Bejger}  \&
  {Lattimer}}{{Haensel} et~al.}{2009}]{haenseletal2009}
{Haensel} P.,  {Zdunik} J.~L.,  {Bejger} M.,   {Lattimer} J.~M.,  2009, \mn@doi
  [\aap] {10.1051/0004-6361/200811605}, \href
  {https://ui.adsabs.harvard.edu/abs/2009A&A...502..605H} {502, 605}

\bibitem[\protect\citeauthoryear{{Haskell} \& {Melatos}}{{Haskell} \&
  {Melatos}}{2015}]{haskellMelatos2015}
{Haskell} B.,  {Melatos} A.,  2015, \mn@doi [International Journal of Modern
  Physics D] {10.1142/S0218271815300086}, \href
  {http://adsabs.harvard.edu/abs/2015IJMPD..2430008H} {24, 1530008}

\bibitem[\protect\citeauthoryear{{Hild}, {Abernathy}, {Acernese},
  {Amaro-Seoane}, {Andersson}, {Arun}, {Barone}  \& et al.}{{Hild}
  et~al.}{2011}]{hildetal2011}
{Hild} S.,  {Abernathy} M.,  {Acernese} F.,  {Amaro-Seoane} P.,  {Andersson}
  N.,  {Arun} K.,  {Barone} F.,   et al. 2011, \mn@doi [Classical and Quantum
  Gravity] {10.1088/0264-9381/28/9/094013}, \href
  {http://adsabs.harvard.edu/abs/2011CQGra..28i4013H} {28, 094013}

\bibitem[\protect\citeauthoryear{{Ho} \& {Lai}}{{Ho} \&
  {Lai}}{2000}]{hoLai2000}
{Ho} W. C.~G.,  {Lai} D.,  2000, \mn@doi [\apj] {10.1086/317085}, \href
  {https://ui.adsabs.harvard.edu/abs/2000ApJ...543..386H} {543, 386}

\bibitem[\protect\citeauthoryear{{Ho}, {Jones}, {Andersson}  \&
  {Espinoza}}{{Ho} et~al.}{2020}]{hoetal2020}
{Ho} W. C.~G.,  {Jones} D.~I.,  {Andersson} N.,   {Espinoza} C.~M.,  2020,
  \mn@doi [\prd] {10.1103/PhysRevD.101.103009}, \href
  {https://ui.adsabs.harvard.edu/abs/2020PhRvD.101j3009H} {101, 103009}

\bibitem[\protect\citeauthoryear{{Horowitz} \& {Kadau}}{{Horowitz} \&
  {Kadau}}{2009}]{horowitzKadau2009}
{Horowitz} C.~J.,  {Kadau} K.,  2009, \mn@doi [\prl]
  {10.1103/PhysRevLett.102.191102}, \href
  {https://ui.adsabs.harvard.edu/abs/2009PhRvL.102s1102H} {102, 191102}

\bibitem[\protect\citeauthoryear{{Ioka}}{{Ioka}}{2001}]{ioka2001}
{Ioka} K.,  2001, \mn@doi [\mnras] {10.1046/j.1365-8711.2001.04756.x}, \href
  {https://ui.adsabs.harvard.edu/abs/2001MNRAS.327..639I} {327, 639}

\bibitem[\protect\citeauthoryear{{Jaranowski}, {Kr{\'o}lak}  \&
  {Schutz}}{{Jaranowski} et~al.}{1998}]{jaranowskiKrolakSchutz1998}
{Jaranowski} P.,  {Kr{\'o}lak} A.,   {Schutz} B.~F.,  1998, \mn@doi [\prd]
  {10.1103/PhysRevD.58.063001}, \href
  {http://adsabs.harvard.edu/abs/1998PhRvD..58f3001J} {58, 063001}

\bibitem[\protect\citeauthoryear{{Jones}}{{Jones}}{2010}]{jones2010}
{Jones} D.~I.,  2010, \mn@doi [\mnras] {10.1111/j.1365-2966.2009.16059.x},
  \href {https://ui.adsabs.harvard.edu/abs/2010MNRAS.402.2503J} {402, 2503}

\bibitem[\protect\citeauthoryear{{Jones}}{{Jones}}{2012}]{jones2012}
{Jones} D.~I.,  2012, \mn@doi [\mnras] {10.1111/j.1365-2966.2011.20238.x},
  \href {https://ui.adsabs.harvard.edu/abs/2012MNRAS.420.2325J} {420, 2325}

\bibitem[\protect\citeauthoryear{{Keer} \& {Jones}}{{Keer} \&
  {Jones}}{2015}]{keerJones2015}
{Keer} L.,  {Jones} D.~I.,  2015, \mn@doi [\mnras] {10.1093/mnras/stu2123},
  \href {https://ui.adsabs.harvard.edu/abs/2015MNRAS.446..865K} {446, 865}

\bibitem[\protect\citeauthoryear{{Kr{\"u}ger}, {Ho}  \&
  {Andersson}}{{Kr{\"u}ger} et~al.}{2015}]{krugerHoAndersson2015}
{Kr{\"u}ger} C.~J.,  {Ho} W.~C.~G.,   {Andersson} N.,  2015, \mn@doi [\prd]
  {10.1103/PhysRevD.92.063009}, \href
  {https://ui.adsabs.harvard.edu/abs/2015PhRvD..92f3009K} {92, 063009}

\bibitem[\protect\citeauthoryear{{Levin} \& {Ushomirsky}}{{Levin} \&
  {Ushomirsky}}{2001}]{levinUshomirsky2001}
{Levin} Y.,  {Ushomirsky} G.,  2001, \mn@doi [\mnras]
  {10.1046/j.1365-8711.2001.04075.x}, \href
  {https://ui.adsabs.harvard.edu/abs/2001MNRAS.322..515L} {322, 515}

\bibitem[\protect\citeauthoryear{{Lower} et~al.,}{{Lower}
  et~al.}{2021}]{loweretal2021}
{Lower} M.~E.,  et~al., 2021, \mn@doi [\mnras] {10.1093/mnras/stab2678}, \href
  {https://ui.adsabs.harvard.edu/abs/2021MNRAS.508.3251L} {508, 3251}

\bibitem[\protect\citeauthoryear{{Lyne} \& {Graham-Smith}}{{Lyne} \&
  {Graham-Smith}}{2012}]{l_g-s_12}
{Lyne} A.,  {Graham-Smith} F.,  2012, {Pulsar Astronomy}.
Cambridge University Press, Cambridge

\bibitem[\protect\citeauthoryear{{Lyne}, {Hobbs}, {Kramer}, {Stairs}  \&
  {Stappers}}{{Lyne} et~al.}{2010}]{lyneetal2010}
{Lyne} A.,  {Hobbs} G.,  {Kramer} M.,  {Stairs} I.,   {Stappers} B.,  2010,
  \mn@doi [Science] {10.1126/science.1186683}, \href
  {https://ui.adsabs.harvard.edu/abs/2010Sci...329..408L} {329, 408}

\bibitem[\protect\citeauthoryear{{Manchester}, {Hobbs}, {Teoh}  \&
  {Hobbs}}{{Manchester} et~al.}{2005}]{manchesteretal2005}
{Manchester} R.~N.,  {Hobbs} G.~B.,  {Teoh} A.,   {Hobbs} M.,  2005, \mn@doi
  [\aj] {10.1086/428488}, \href
  {http://adsabs.harvard.edu/abs/2005AJ....129.1993M} {129, 1993}

\bibitem[\protect\citeauthoryear{{McDermott}, {van Horn}  \&
  {Hansen}}{{McDermott} et~al.}{1988}]{mcdermottVanHornHansen1988}
{McDermott} P.~N.,  {van Horn} H.~M.,   {Hansen} C.~J.,  1988, \mn@doi [\apj]
  {10.1086/166044}, \href
  {https://ui.adsabs.harvard.edu/abs/1988ApJ...325..725M} {325, 725}

\bibitem[\protect\citeauthoryear{{Melatos} \& {Link}}{{Melatos} \&
  {Link}}{2014}]{melatosLink2014}
{Melatos} A.,  {Link} B.,  2014, \mn@doi [\mnras] {10.1093/mnras/stt1828},
  \href {https://ui.adsabs.harvard.edu/abs/2014MNRAS.437...21M} {437, 21}

\bibitem[\protect\citeauthoryear{{Ogata} \& {Ichimaru}}{{Ogata} \&
  {Ichimaru}}{1990}]{ogataIchimaru1990}
{Ogata} S.,  {Ichimaru} S.,  1990, \mn@doi [\pra] {10.1103/PhysRevA.42.4867},
  \href {https://ui.adsabs.harvard.edu/abs/1990PhRvA..42.4867O} {42, 4867}

\bibitem[\protect\citeauthoryear{{Owen}}{{Owen}}{2010}]{owen2010}
{Owen} B.~J.,  2010, \mn@doi [\prd] {10.1103/PhysRevD.82.104002}, \href
  {https://ui.adsabs.harvard.edu/abs/2010PhRvD..82j4002O} {82, 104002}

\bibitem[\protect\citeauthoryear{{Owen}, {Lindblom}, {Cutler}, {Schutz},
  {Vecchio}  \& {Andersson}}{{Owen} et~al.}{1998}]{owenetal1998}
{Owen} B.~J.,  {Lindblom} L.,  {Cutler} C.,  {Schutz} B.~F.,  {Vecchio} A.,
  {Andersson} N.,  1998, \mn@doi [\prd] {10.1103/PhysRevD.58.084020}, \href
  {https://ui.adsabs.harvard.edu/abs/1998PhRvD..58h4020O} {58, 084020}

\bibitem[\protect\citeauthoryear{{Palfreyman}, {Dickey}, {Hotan}, {Ellingsen}
  \& {van Straten}}{{Palfreyman} et~al.}{2018}]{palfreymanetal2018}
{Palfreyman} J.,  {Dickey} J.~M.,  {Hotan} A.,  {Ellingsen} S.,   {van Straten}
  W.,  2018, \mn@doi [\nat] {10.1038/s41586-018-0001-x}, \href
  {https://ui.adsabs.harvard.edu/abs/2018Natur.556..219P} {556, 219}

\bibitem[\protect\citeauthoryear{{Reitze} et~al.,}{{Reitze}
  et~al.}{2019}]{reitzeetal2019}
{Reitze} D.,  et~al., 2019, in Bulletin of the American Astronomical Society.
  p.~35 (\mn@eprint {arXiv} {1907.04833})

\bibitem[\protect\citeauthoryear{{Ruderman}}{{Ruderman}}{1969}]{ruderman1969}
{Ruderman} M.,  1969, Nature, 223, 597

\bibitem[\protect\citeauthoryear{{Shaw} et~al.,}{{Shaw}
  et~al.}{2018}]{shawetal2018}
{Shaw} B.,  et~al., 2018, \mn@doi [\mnras] {10.1093/mnras/sty1294}, \href
  {http://adsabs.harvard.edu/abs/2018MNRAS.478.3832S} {478, 3832}

\bibitem[\protect\citeauthoryear{{Sidery}, {Passamonti}  \&
  {Andersson}}{{Sidery} et~al.}{2010}]{sideryPassamontiAndersson2010}
{Sidery} T.,  {Passamonti} A.,   {Andersson} N.,  2010, \mn@doi [\mnras]
  {10.1111/j.1365-2966.2010.16497.x}, \href
  {https://ui.adsabs.harvard.edu/abs/2010MNRAS.405.1061S} {405, 1061}

\bibitem[\protect\citeauthoryear{{Strohmayer}, {Ogata}, {Iyetomi}, {Ichimaru}
  \& {van Horn}}{{Strohmayer} et~al.}{1991}]{strohmayeretal1991}
{Strohmayer} T.,  {Ogata} S.,  {Iyetomi} H.,  {Ichimaru} S.,   {van Horn}
  H.~M.,  1991, \mn@doi [\apj] {10.1086/170231}, \href
  {https://ui.adsabs.harvard.edu/abs/1991ApJ...375..679S} {375, 679}

\bibitem[\protect\citeauthoryear{{Tenorio}, {Modafferi}, {Keitel}  \&
  {Sintes}}{{Tenorio} et~al.}{2022}]{tenorioetal2022}
{Tenorio} R.,  {Modafferi} L.~M.,  {Keitel} D.,   {Sintes} A.~M.,  2022,
  \mn@doi [\prd] {10.1103/PhysRevD.105.044029}, \href
  {https://ui.adsabs.harvard.edu/abs/2022PhRvD.105d4029T} {105, 044029}

\bibitem[\protect\citeauthoryear{{The LIGO Scientific Collaboration}
  et~al.,}{{The LIGO Scientific Collaboration} et~al.}{2021}]{LVK2021GWTC3}
{The LIGO Scientific Collaboration} et~al., 2021, arXiv e-prints, \href
  {https://ui.adsabs.harvard.edu/abs/2021arXiv211103606T} {p. arXiv:2111.03606}

\bibitem[\protect\citeauthoryear{{The LIGO Scientific Collaboration}
  et~al.,}{{The LIGO Scientific Collaboration}
  et~al.}{2022}]{LVK2022magnetarbursts}
{The LIGO Scientific Collaboration} et~al., 2022, arXiv e-prints, \href
  {https://ui.adsabs.harvard.edu/abs/arXiv:2210.10931} {p. arXiv:2210.10931}

\bibitem[\protect\citeauthoryear{{Thorne}}{{Thorne}}{1980}]{thorne1980}
{Thorne} K.~S.,  1980, \mn@doi [Reviews of Modern Physics]
  {10.1103/RevModPhys.52.299}, \href
  {https://ui.adsabs.harvard.edu/abs/1980RvMP...52..299T} {52, 299}

\bibitem[\protect\citeauthoryear{{Walsh} et~al.,}{{Walsh}
  et~al.}{2016}]{walshetal2016}
{Walsh} S.,  et~al., 2016, \mn@doi [\prd] {10.1103/PhysRevD.94.124010}, \href
  {http://adsabs.harvard.edu/abs/2016PhRvD..94l4010W} {94, 124010}

\bibitem[\protect\citeauthoryear{{Wen}, {Li}, {Chen}  \& {Zhang}}{{Wen}
  et~al.}{2019}]{wenetal2019}
{Wen} D.-H.,  {Li} B.-A.,  {Chen} H.-Y.,   {Zhang} N.-B.,  2019, \mn@doi [\prc]
  {10.1103/PhysRevC.99.045806}, \href
  {https://ui.adsabs.harvard.edu/abs/2019PhRvC..99d5806W} {99, 045806}

\bibitem[\protect\citeauthoryear{{Yim} \& {Jones}}{{Yim} \&
  {Jones}}{2022}]{yimJones2022}
{Yim} G.,  {Jones} D.~I.,  2022, \mn@doi [\mnras] {10.1093/mnras/stac009},
  \href {https://ui.adsabs.harvard.edu/abs/2022MNRAS.511.1942Y} {511, 1942}

\bibitem[\protect\citeauthoryear{{Yu} et~al.,}{{Yu} et~al.}{2013}]{yuetal2013}
{Yu} M.,  et~al., 2013, \mn@doi [\mnras] {10.1093/mnras/sts366}, \href
  {https://ui.adsabs.harvard.edu/abs/2013MNRAS.429..688Y} {429, 688}

\makeatother
\end{thebibliography}







\bsp	
\label{lastpage}
\end{document}